\documentclass[aps,amsfonts,pra,preprint,singlecolumn,showpacs]{revtex4}
\usepackage{hyperref}
\usepackage{graphicx}
\usepackage{color}

\usepackage{amsmath}

\def\NOT(#1,#2){\OneQubitGate(#1,#2){$X$}}

\begin{document}

\title{Experimental quantum error correction with
high fidelity}
\author{
Jingfu Zhang$^1$,  Dorian Gangloff $^{1*}$, Osama Moussa$^1$,
  and Raymond Laflamme$^{1,2}$\\
\it {$^1$Institute for Quantum Computing and Department of
Physics,
University of Waterloo, Waterloo, Ontario, Canada N2L 3G1\\
$^2$Perimeter Institute for Theoretical Physics, Waterloo, Ontario, N2J 2W9,
Canada\\
}}
\date{\today}

\begin{abstract}
 More than ten years ago a first step towards quantum
error correction (QEC) was implemented [Phys. Rev. Lett. 81, 2152
(1998)]. The work showed there was sufficient control in nuclear
magnetic resonance (NMR) to implement QEC, and demonstrated that
the error rate changed from $\epsilon$ to approximatively
$\sim\epsilon^2$.  In the current work we reproduce a similar
experiment using control techniques that have been since
developed, such as GRAPE pulses. We show that the fidelity of the
QEC gate sequence, and the comparative advantage of QEC are
appreciably improved. This advantage is maintained despite the
errors introduced by the additional operations needed to protect
the quantum states.
\end{abstract}
 \maketitle

\section{Introduction}
Quantum computers could solve some problems faster than classical
computers \cite{NielsenChuang}. Performing a quantum computation
relies on the ability to preserve the coherence of quantum states
long enough for gates composing the algorithm to be implemented.
In practice, the quantum coherence is sensitive to the
uncontrolled environment and easily damaged by the interactions
with the environment, a process called decoherence
\cite{phystoday}. To protect the fragile quantum coherence  needed
for quantum computation, schemes of quantum error correction (QEC)
and fault- tolerant quantum computation have been developed
\cite{knill1998}.

The 3-bit QEC code was implemented in a liquid-state NMR quantum
information processor in 1998 as the first experimental
demonstration of QEC \cite{QEC98}. More recently, it has been
implemented in trapped ion and solid-state systems
\cite{ion,solidNMR}. Here we report on using the GRAPE algorithm
\cite{grape} to implement a high fidelity version of the 3-bit QEC
code for phase errors in liquid state NMR. The errors due to
natural transversal relaxation are shown to be suppressed to a
first order. In comparison with the work performed in 1998
\cite{QEC98}, the pulse sequence fidelity is improved by about
$20\%$, and the reduction of the first order in the decay of the
remaining polarization after error correction is improved by a
factor of $\sim 2.3$. The advantage of the QEC is obtained
although the extra operations for protecting the quantum states in
QEC are subject to errors in implementation.

\section{Experimental procedure and results}
\begin{figure}
\includegraphics[width=4in]{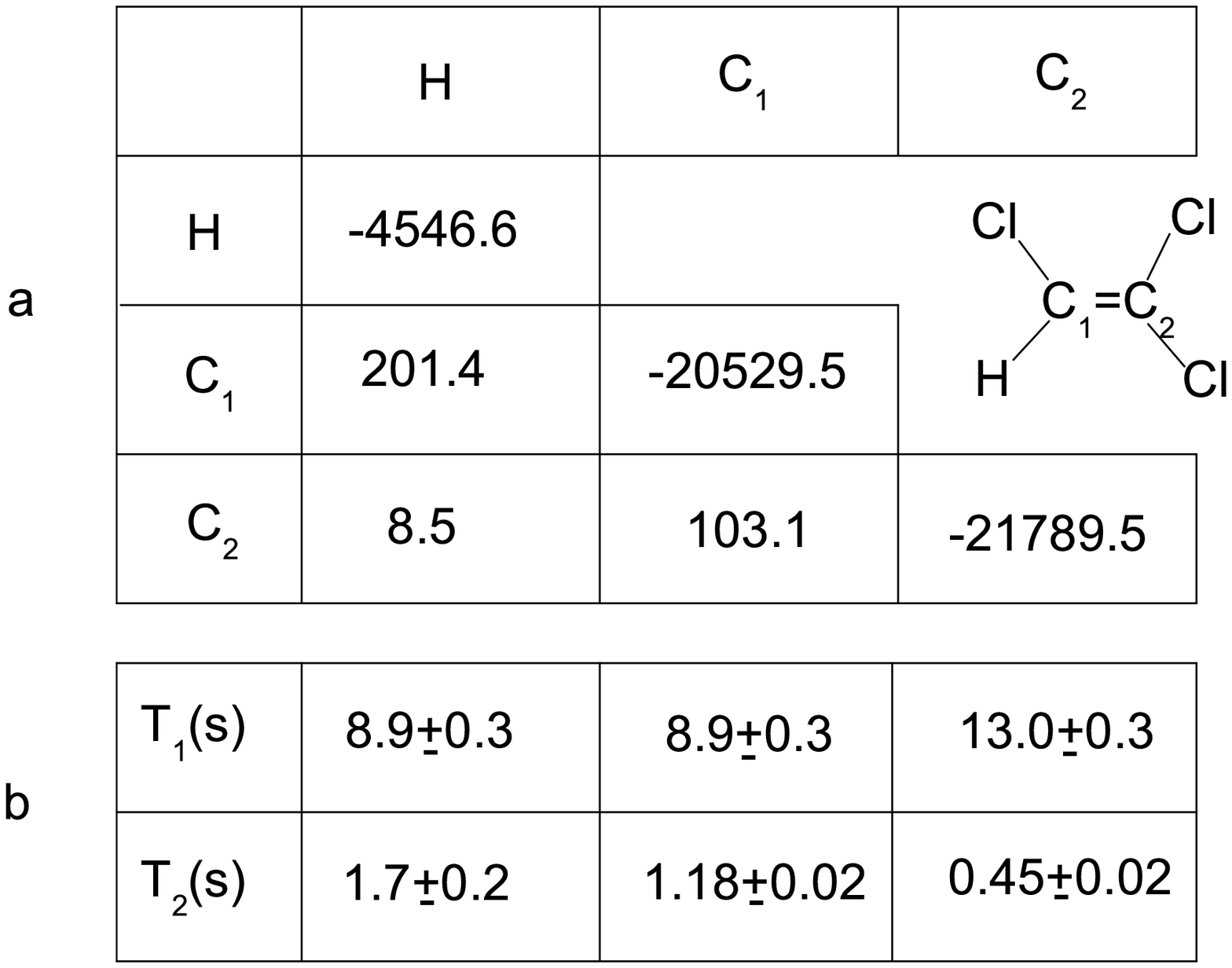} %
\caption{ Parameters of the spin qubits. (a) Chemical shifts shown
as the diagonal terms and the couplings  between spins shown as
the non-diagonal terms in Hz. The inset shows the molecule
structure where the three qubits are H, C$_1$ and C$_2$. (b) The
relaxation times $T_1$'s are measured by the standard inversion
recovery sequence. $T_2$'s are measured by the Hahn-echo with one
refocusing pulse, by ignoring the strong coupling in the
Hamiltonian (\ref{ham}).} \label{figmol}
\end{figure}

  In the current implementation, we use $^{13}$C labelled
trichloroethylene (TCE) dissolved in d-chloroform as the sample.
Data were taken with a Bruker DRX 700 MHz spectrometer. The
structure of the molecule and the parameters of the spin qubits
are shown in Fig. \ref{figmol}, where we denote H as qubit 1,
C$_{1}$  as qubit 2  and C$_{2}$  as qubit 3. The Hamiltonian of
the three-spin system can be written as
\begin{equation}\label{ham}
    H=-\pi\sum_{i=1}^3 \nu_{i}Z_{i}
  +\frac{\pi}{2} [J_{12}Z_{1}Z_{2}
  + J_{13}Z_{1}Z_{3}+
  J_{23}(X_{2}X_{3}+Y_{2}Y_{3}+Z_{2}Z_{3})]
\end{equation}
where $X_{i}$, $Y_{i}$, $Z_{i}$ denote the Pauli matrices with $i$
indicating the spin location, $\nu_{i}$ denotes the chemical shift
of spin $i$, and $J_{ij}$ denotes the spin-coupling between spins
$i$ and $j$. The two carbon spins are treated in the strongly
coupled regime, because the difference in frequencies between the
two carbons is not large enough for the weak coupling
approximation \cite{NMRrew}.

We exploit radio-frequency (r.f.) spin selection techniques to
improve the linewidth, and hence the coherence, of the ensemble
qubits \cite{image,knill}. The effect of pulse imperfections due
to r.f. inhomogeneities is reduced by spatially selecting
molecules from a small region in the sample through  the r.f.
power. We choose C$_1$ as the qubit to carry the state for
encoding and the output state after decoding and error correction.
  The labelled pseudo-pure states $\mathbf{0}X\mathbf{0}$ and $\mathbf{0}Y\mathbf{0}$, used as the reference states with blank ancilla,
   are prepared by the circuit in Ref. \cite{knill}, where the  order is arranged as  qubits $1$ to $3$
   and $\mathbf{0}\equiv |0\rangle\langle0|$.
The qubit readout is performed on C$_{1}$, and the signals are
normalized with respect to $\mathbf{0}X\mathbf{0}$ or
$\mathbf{0}Y\mathbf{0}$, for different input states.

\begin{figure}
\includegraphics[width=4in]{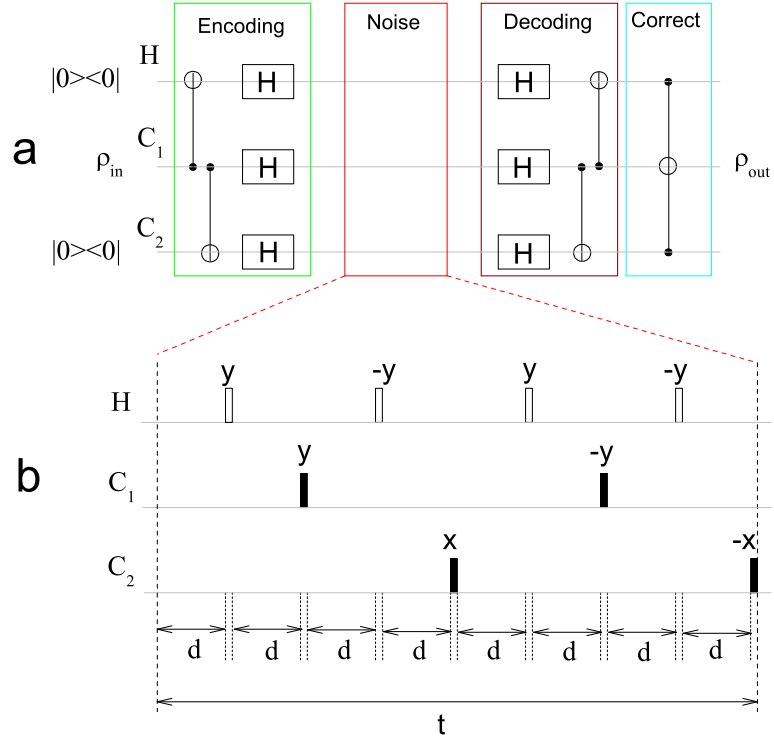} 
\caption{(Color online).   Quantum  network for quantum error
correction (a) where the $T_2$ noise is introduced by a variable
time  delay implemented by the pulse sequence (b) which refocuses
the evolution of the Hamiltonian (\ref{ham}) to an identity
operation with theoretical fidelity higher than $99.96\%$. In (b)
the unfilled rectangle represents a hard $\pi$ pulse with duration
of $20$ $\mu$s. The filled rectangle represents a GRAPE $\pi$
pulse selective for C$_1$ or C$_2$ with duration of $2$ ms. The
phases of the pulses are denoted above the rectangles.}
\label{figcir}
\end{figure}

The quantum network used for implementing the QEC code is shown as
Fig. \ref{figcir} (a), where $\rho_{in}$ is chosen as $X$, $Y$ and
$Z$, in separate sequences. We optimize the encoding operation,
and the decoding operation combined with the error correction as
two GRAPE pulses \cite{grape} with theoretical fidelity $>99.9\%$.
To test the ability of the code to correct for the natural
dephasing errors due to the transversal relaxation of the spins,
the internal spin Hamiltonian (\ref{ham}) is refocused during the
time delay implemented between the encoding and decoding
processes. The refocusing pulse sequence is shown in Fig.
\ref{figcir} (b) where the selective $\pi$ pulses applied to spin
H are hard rectangle pulses with a duration of $20$ $\mu$s, while
the
 $\pi$ pulses applied to C$_1$ or C$_2$ are GRAPE
pulses with a duration of $2$ ms. Taking into account the strong
coupling in the Hamiltonian  (\ref{ham}), we choose the phases of
the $\pi$ pulses shown in Fig. \ref{figcir} (b) to obtain a
fidelity $|Tr\{U_{refocus}E\}|/8>99.96\%$, where $U_{refocus}$
denotes the simulated unitary implemented by the pulse sequence,
and $E$ denotes the identity operation.

We choose the input states as $\rho_{in}=X$, $Y$ and $Z$, and
measure the polarization that remains after error correction in
$\rho_{out}$. The polarization ratios are denoted as $f_x$, $f_y$
and $f_z$. We use ''entanglement fidelity", represented as
\begin{equation}\label{entfid}
   f = (1+f_x+f_y+f_z)/4
\end{equation}
to characterize how well the quantum information in $\rho_{in}$ is
preserved \cite{entfid}.

\begin{figure}
\includegraphics[width=6in]{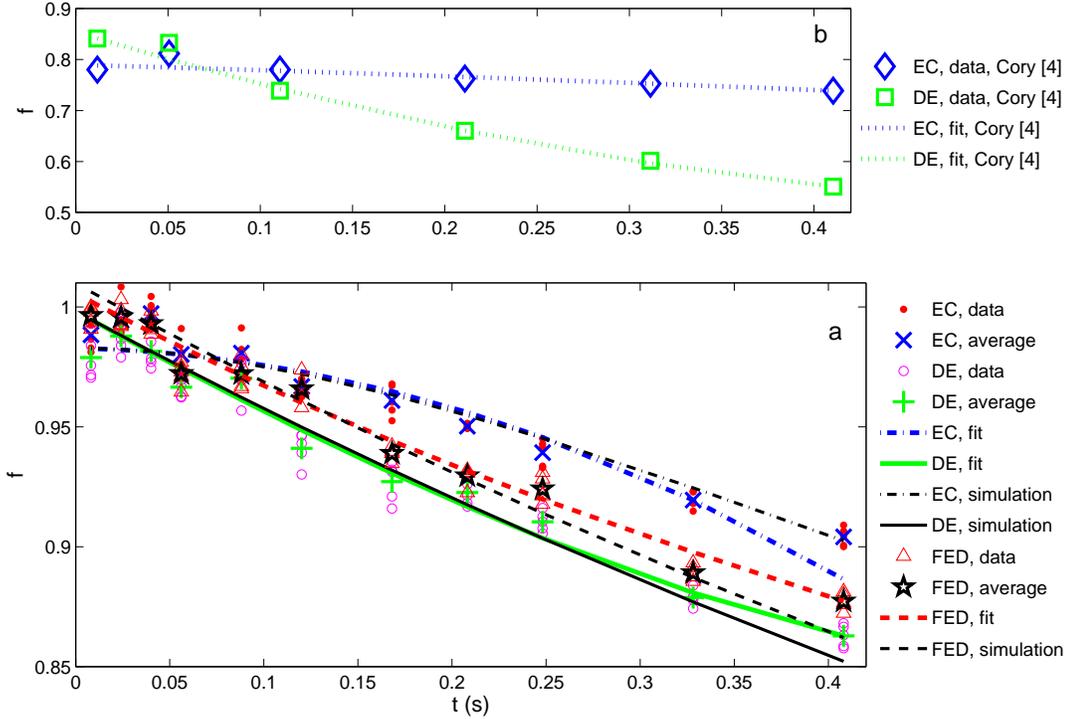} 
\caption{(Color online). (a) Experimental results for error
correction (EC), decoding (DE), and free evolution decay (FED).
For each delay time, we take five data points by repeating
experiments, shown as {\large$\bullet$} for EC, $\bigcirc$ for DE
and $\triangle$ for FED. The averages are shown as ''$\times$'',
''+'' and {\large $\star$}, which can be fitted as $0.9828 -0.0166
t -0.5380t^2
 +0.0014t^3$  with relative fitting error $0.73\%$,
$0.9982 -0.4361 t + 0.1679t^2+ 0.2152t^3$ with relative fitting
error $0.57\%$ and $1.0056 -0.4164 t + 0.3363t^2 - 0.2123t^3$ with
relative fitting error $0.45\%$, shown as the thick dash-dotted,
solid and dashed curves, respectively. The ratios of the
first-order decay terms in the fitted curves are calculated as
$26.2\pm 0.3$ for DE and EC, and $25.0\pm 0.3$ for FED and EC,
respectively. The thin dash-dotted, solid and dashed curves show
the fitting results using the ideal data points from simulation by
introducing factors of $0.983\pm0.006$, $0.998\pm0.007$ and
$1.0098\pm 0.0064$ for EC, DE and FED, respectively. (b) Results
in the previous experiment \cite{QEC98}, shown as the data marked
by ''$\diamond$" and ''$\square$" for EC and DE, which can be
fitted as $0.7895 -0.0957 t -0.0828 t^2 +0.0370t^3$ and $0.8539
-1.1021 t + 0.8696t^2 +
 0.0378t^3$ with relative fitting errors $0.89\%$ and $0.98\%$, respectively.
 The ratio of the first-order decay terms is $11.5\pm 0.2$.}
\label{figres}
\end{figure}

   The experimental results for QEC are shown in Fig. \ref{figres} (a).
For each delay time, five experiments are repeated in order to
average the random experimental errors in implementation. The
results of error correction (EC) are represented by
{\large$\bullet$}. By averaging the points for each delay time, we
obtain the averaged entanglement fidelity $f$ shown as $\times$,
which can be fitted to $0.9828 -0.0166 t -0.5380t^2
 +0.0014t^3$, with relative fitting error $0.73\%$, shown as the thick dash-dotted curve.

  In order to estimate the performance of the error correction for the encoded states,
we calculate the entanglement fidelity of decoding (DE) through
measuring the remaining polarization before the application of the
Toffoli gate, used as the error-correcting step. In this case, the
decoding operation is implemented by one GRAPE pulse with
theoretical fidelity $>99.9\%$. Similar to the measurement for
error correction, we also repeat five experiments for each delay
time. The results are shown as $\bigcirc$ in Fig. \ref{figres}
(a), and the data points after average are marked by +, which can
be fitted as $0.9982 -0.4361 t + 0.1679t^2+ 0.2152t^3$, with
relative fitting error $0.57\%$, shown as the thick solid curve.
Here the ratio of the first order decay terms for the two fits is
found to be $26.2\pm 0.3$. The important reduction of the first
order decay term indicates the high quality of state stabilization
provided by QEC. As a comparison, we include the experimental data
from Ref. \cite{QEC98}, which are marked as $\diamond$ and
$\square$ in Fig. \ref{figres} (b) for the results of QEC and
decoding. The data can be fitted as $0.7895 -0.0957 t -0.0828 t^2
+0.0370t^3$ and $0.8539 -1.1021 t + 0.8696t^2 +
 0.0378t^3$ with relative fitting errors $0.89\%$ and $0.98\%$, respectively.
 The ratio of the first-order decay terms is $11.5\pm 0.2$.

 In implementing the QEC code, the operations
associated with encoding, decoding and error correction are
subject to errors, which would lower the ability of the code to
protect the quantum states. To estimate the effects of the errors,
we measure the free evolution decay (FED) of $\rho_{in}$ under the
refocusing sequence shown in Fig. \ref{figcir} (b). Five
experiments are repeated for each delay time, and the experimental
data for $f$ are shown as $\triangle$ in Fig. \ref{figres} (a).
The averaging points, shown as {\large $\star$}, can be fitted as
$1.0056 -0.4164 t + 0.3363t^2 - 0.2123t^3$ with relative fitting
error $0.45\%$, shown as the dashed curve. The ratio of the first
order decay terms in the fits of FED  and EC is $25.0\pm 0.3$.
Through comparing the results of QEC and FED, one can find that
the errors removed by the QEC code can exceed the errors
introduced by the extra operations required by the code for delay
time $>0.0672$ s ($\sim 6\%$ of C$_{1}$'s $T_2$).

\section{Discussion}
The pulse durations for encoding, decoding, and the combination of
decoding and error correction are 8 ms, 8ms, and 13.6 ms,
respectively. We exploit the results from simulation with ideal
pulses to estimate the errors due to the imperfection in pulse
implementation. In simulation, we choose an uncorrelated error
model for  $T_{2}$ errors and ignore $T_{1}$ errors \cite{cory98}.
We represent the measured fidelity as $f = Af_{ideal}$, where
$f_{ideal}$ denotes the ideal fidelity by simulation and $A$
denotes a factor to estimate the deviation between experiment and
simulation. One should note that the theoretical entanglement
fidelity of DE is the same as FED \cite{cory98}. By fitting the
data, we obtain $A = 0.983\pm0.006$, $0.998\pm0.007$ and
$1.0098\pm 0.0064$ for EC, DE and FED, respectively. The fitting
results are shown as the thin
dash-dotted, solid and dashed curves in Fig. \ref{figres} (a). 
From the simulation results, we estimate the errors in
implementing the operations associated with the QEC codes are
about $1.2\%$ for DE and $2.7\%$ for EC.


\section{Conclusion}
We optimize the encoding, decoding, and error correction as GRAPE
pulses with high theoretical fidelities ($>99.9\%$). The
refocusing sequence is exploited to suspend the evolution of the
Hamiltonian (\ref{ham}) with high fidelity ($>99.96\%$). The
quality of readout signals are further improved by  r.f.
selection. Compared with the experimental results of QEC obtained
in 1998 \cite{QEC98}, the pulse sequence fidelity is improved by
about $20\%$.  By the comparison with the free evolution decay,
one can benefit from QEC even when errors exist in implementing
the operations required for QEC. The improvement provided by the
error correction is also demonstrated by the reduction of the
first order in the decay of the remaining polarization after error
correction, compared with the decay of the encoded states
recovered by decoding and free evolution decay of the input
states.  The ``QEC advantage'' for the encoded states is improved
by a factor of $\sim 2.3$ from the 1998 result. In the current
experiment, the second order term in the decay after error
correction is larger than the previous experiment, because of the
larger phase errors due to the shorter $T_2$ time constants [see
Fig. \ref{figmol} (b)], noting that $T_2$'s are $3$ s for H, $1.1$
s for C$_1$ and $0.6$ s for C$_2$ in the previous experiment
\cite{QEC98}.
The experimental errors arise mainly from the imperfection in
implementing the GRAPE pulses. Additionally inhomogeneities of
magnetic fields and the limitation of $T_{1}$ also contribute to
errors.

\section{Acknowledgments}
The authors acknowledge professor D. G. Cory  for helpful
discussions.\\

$^{*}$ Current address: Department of Physics, Massachusetts
Institute of Technology, Cambridge, MA, USA 02139

\end{document}